# Spin-dependent Seebeck effect in non-local spin valve devices

Mikhail Erekhinsky,[1] Fèlix Casanova,[2,3] Ivan K. Schuller,[1] and Amos Sharoni[1,4]

[1] Center for Advanced Nanoscience, Department of Physics, University of California - San Diego, La Jolla, California 92093, USA
[2] CIC nanoGUNE, 20018 Donostia-San Sebastian, Basque Country
[3] IKERBASQUE, Basque Foundation for Science, 48011 Bilbao, Basque Country
[4] Department of Physics, Institute of Nanotechnology and Advanced Materials, Bar Ilan University, Ramat-Gan 52900, Israel

**We performed measurements of Py/Cu and Py/Ag lateral spin valves as function of injection current direction and magnitude. Above a "critical" current, there is an unexpected dependence of spin injection on current direction. Positive currents show higher polarization of spin injection than negative. This implies that in addition to current-induced spin injection, there is a thermally induced injection from a spin-dependent Seebeck effect. A temperature gradient in the Py electrode, caused by Joule heating, is responsible for injecting excess spins into the non-magnetic channel. This effect has important consequences for understanding high-current spin-based devices, such as spin transfer torque devices.**

Thermoelectric properties are of major importance in many nano-electronic and spintronic applications.[1-3] However, the importance of thermal effects has been largely overlooked in spintronics devices. Recently, the interactions between electrical, spin and heat (or entropy) transport have received considerable attention and are all lumped in the term "spin-caloritronics". Examples include thermal spin transfer-torque,[4] spin Nernst,[5] spin Peltier [6] and spin Seebeck [7-9] effects.

Spin-caloritronics may have an important role in future spintronic devices. While thermal effects can be used to generate new functionalities,[9, 10] they might have deleterious effects on device performance. For example, in many applications needing high current densities (e.g., spin-torque-induced switching in magnetic random access memory), spurious thermal gradients may arise.[11] Thus, an understanding of these effects has important implications for spin based devices.

We found an unexpected asymmetry in non-local ferromagnetic (FM)/ non-magnetic (NM) metallic lateral spin valves.[12] At high current densities (~ $4.5 \times 10^{11}$ A/m$^2$ in the FM) the spin injection is larger when the charge current flows from the NM into the FM, compared to spin extraction when the charge current flows from the FM into the NM. We propose that this asymmetry arises from spin-caloritronics: local Joule heating produces a temperature gradient in the FM near the FM/NM interface, which induces additional thermal spin injection, due to the spin-dependent Seebeck effect. The results are in semi quantitative agreement with experimental data of the spin-dependent Seebeck coefficient; Peltier or spin blockade effects are ruled out.

Lateral spin valves were fabricated by a two-angle shadow evaporation method,[11] providing a clean interface between FM electrodes made of Py and the NM channel made of Cu or Ag. Such interfaces are necessary due to the high current densities we apply. We note that in this geometry the bare injecting FM electrode is the thinnest part of the device (see inset of Fig. 1a and cartoon in Fig. 4). Therefore, it has the highest current density and is responsible for most of the Joule heating in the device.

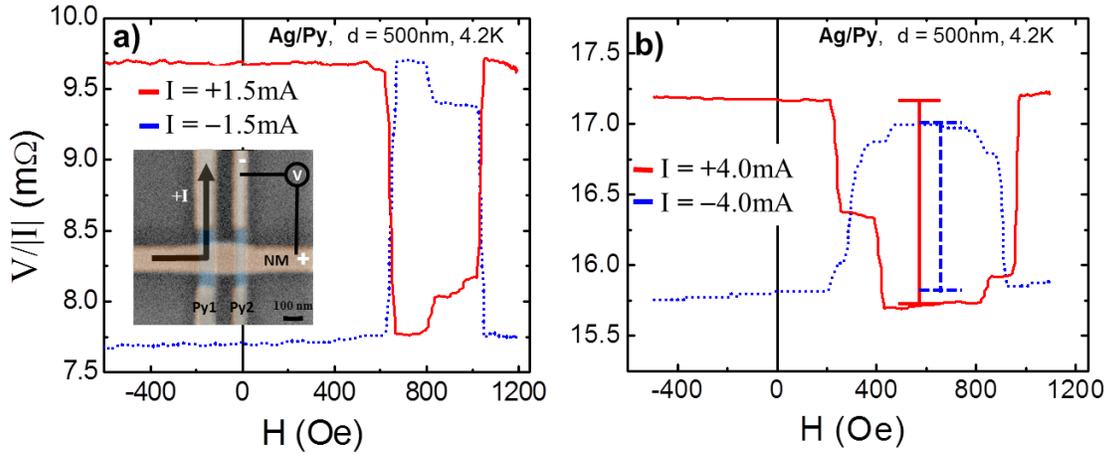

FIG. 1. Non-local voltage normalized to the current magnitude vs. magnetic field for positive (red solid) and negative (dotted blue) currents, measured for an Ag/Py device with $d$=500 nm at $T$=4.2 K. **(a)** $I$=1.5 mA, NLSV signals are the same. **(b)** $I$= 4 mA, NLSV signal for positive current (red solid bar) is larger than that for negative current (blue dotted bar). Inset in **(a)** Scanning electron microscope image of the device. Py electrodes (blue) and NM bar (colored orange) are shown. Notice areas of overlap and small area where only Py exists. Current direction and voltage setup are marked.

Samples with 6-8 lateral spin valves each in close physical proximity were prepared under identical deposition conditions to avoid possible systematic errors.[13] Each spin valve device consists of two Py electrodes crossed by a common Cu or Ag channel, as shown in the inset of Fig. 1(a). In all devices the thickness of the Py electrodes is 35 nm and the edge-to-edge distance ($d$) between them is varied from 200 to 2000 nm. In each device, one of the Py electrodes is 100 nm wide and the other is 150 nm providing separate control over the magnetization of each electrode. [14, 15] The thicknesses of the Cu and Ag channels are 120 nm and 140 nm respectively, and their width is fixed at 250 nm.

Non-local electrical measurements [12] were performed in a Helium-flow cryostat at 4.2 K as a function of external magnetic field, using a dc current source and a nano-voltmeter. Two different measurement techniques were used. The "dc reversal" technique is similar to the standard ac lock-in measurement and cannot distinguish between positive and negative currents. In the "pure dc" technique the current is applied with one polarity and the voltage is compared to the one measured with zero applied current to remove any parasitic capacitance, offsets in the voltmeter and to reduce noise. [11] The positive current direction is defined when it flows from the NM (horizontally aligned bar, see inset in Fig. 1a) into the FM (vertically aligned electrode). When the external magnetic field is swept, the relative magnetic orientation of the FM electrodes changes from parallel to antiparallel and to parallel again. Different magnetic orientations produce voltage changes across the FM/NM detector. Thus, for positive (negative) currents, the voltage changes from a high (low) value for parallel magnetization orientation to a low (high) value for antiparallel.[11] The difference between the high and low voltages normalized to the current magnitude is denoted as the Non-Local Spin Valve (NLSV) signal, which is proportional to the spin accumulation at the detector. Figure 1 shows the non-local voltage normalized to the current magnitude ($V/|I|$) as function of magnetic field for a Py/Ag sample with an electrode spacing d = 500 nm, measured by "pure dc" technique. Figure 1a shows $V/|I|$ for positive and negative excitation currents of 1.5 mA. The NLSV signal is inverted for negative current (parallel state has low resistance and antiparallel state high), but has the same magnitude (~1.9 mΩ). In addition, both have the same ~8.8 mΩ offset. The inversion is caused by the production of opposite spin population (spin injection vs. spin extraction) under the FM injector.[11] Joule heating at the FM injector produces a temperature gradient and a consequent dc thermoelectric signal at the FM detector electrode, responsible for the offset.[11, 16] We note that the mid state jumps appearing around 900 Oe are due to partial flipping of the FM injector or detector, and have been observed in other NLSV measurements.[17, 18]

Figure 1b shows the same measurement as in Fig. 1a, but for currents of 4.0 mA. Note that this figure has the same 2.5 mΩ y-axis range as Fig. 1a, but a different offset. The increased offset from 8.8 mΩ to 16.5 mΩ and the decreased NLSV are a consequence of the larger Joule heating at higher current densities.[11, 16, 19] Although the general curve-shapes are similar to Fig. 1a, there now is a striking difference in the NLSV signal between positive and negative currents that cannot be assigned to simple Joule heating.

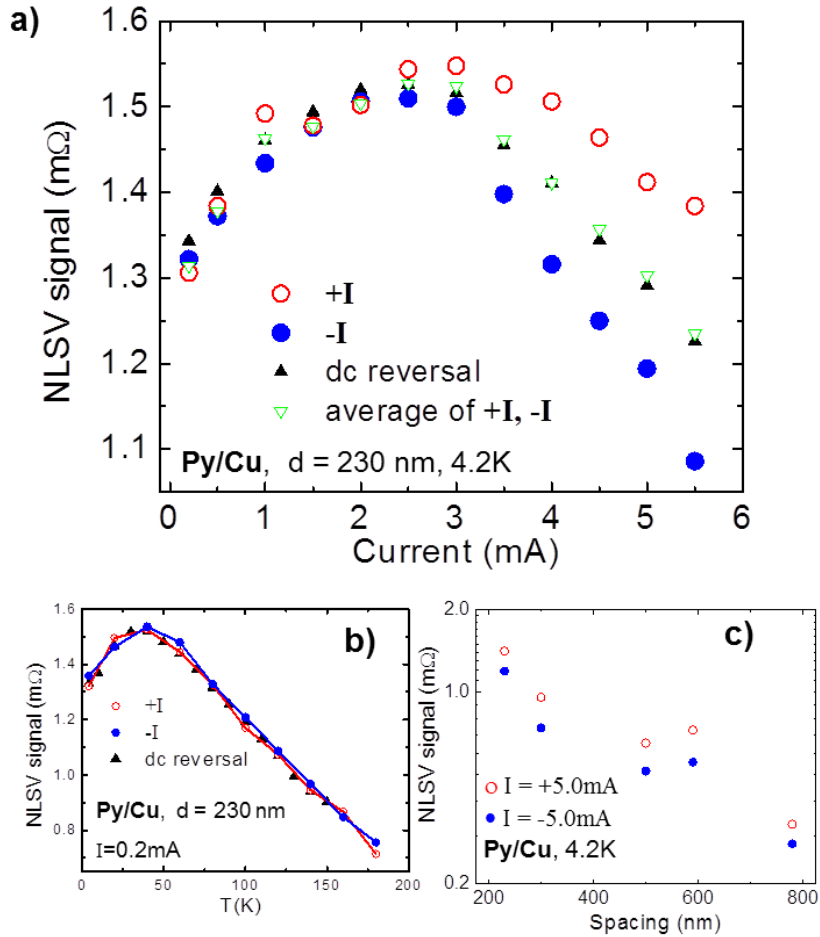

FIG. 2. **(a)** NLSV signal vs. current measured for a Py/Cu device with $d$ = 230 nm at $T$=4.2K. Measurements are done with positive currents (red empty circles), negative currents (blue full circles) and dc-reversal (black triangles). Green empty triangles are the averages of positive and negative current measurements. **(b)** NLSV signal vs. temperature of the same device measured at $I$= 0.2 mA, for positive currents, negative currents and dc-reversal. **(c)** NLSV signal vs. FM electrode spacing d (same sample), measured for $I$=+/- 5 mA. The y-axis is logarithmic scale.

Figure 2a shows the NLSV signal as a function of the injection current up to 5.5 mA, in a Py/Cu device with d=230 nm. The "dc reversal" NLSV equals the average of the positive and negative current measurements, as expected. The overall increase and decrease of the NLSV signal vs. current curve (Fig. 2a) is similar to its temperature dependence [19] at low (0.2 mA) currents (Fig. 2b) as expected from Joule heating.[11] Moreover, positive currents, negative currents and "dc reversal" measurement at low currents give the same results as a function of Temperature. Thus the overall shape of the NLSV signal (in Fig. 2a) originates from the heating of the device. However, starting at ~2.5 mA, a separation appears in the NLSV signal for different current directions. This difference in the NLSV signal for positive and negative currents cannot be merely assigned to a temperature dependence of the device behavior, as shown in Fig 2b.

An important issue at this stage is whether the differences in the NLSV signal (Fig 2a) between positive (+*I*) and negative (−*I*) current arises from the simple Peltier effect. The Peltier coefficient (*Π*) of Py is negative whereas for Cu and Ag it is positive.[8, 20, 21] Thus, when current flows from Cu (or Ag) into Py (+*I*) heat flows towards the interface in the Cu (or Ag) and also towards the interface in the Py. In this case the Peltier effect tends to heat the interface. The opposite will be true for –*I* and therefore the interface will be cooled by the Peltier effect. Above the peak (>50 K) the NLSV signal decreases with increasing temperature as shown in Fig 2b. This implies that when the interface is hotter (+*I*), the NLSV signal (open symbols in Fig. 2a) should be smaller than for the cooler interface produced by –*I* (solid symbols in Fig 2b). This unequivocally rules out a simple Peltier effect as a source of the differences and therefore this must be connected to spin diffusion across the interface.

Figure 2c shows the NLSV signal as a function of the electrode spacing for a Py/Cu sample for two different currents *I*= +5 mA (open red circles) and *I*= −5 mA (full blue circles), on a semi-log plot. For the two currents, the NLSV signal decreases with increasing distance between electrodes. On this semi-log plot the offset between the signal for positive and negative currents remains approximately constant. This indicates that the spin diffusion lengths for the two currents are the same, and therefore the difference arises from the effective spin injection efficiencies. To extract the NM spin diffusion length ($\lambda_{NM}$) and the FM effective spin polarization ($\alpha_{FM}$)[15, 22] for different injection currents, we measured the NLSV signal as a function of the FM electrodes spacing. [13, 15, 22] Applying (as is traditional) the one-dimensional spin diffusion model with transparent interfaces to our non-local geometry, the NLSV signal as a function of $\lambda_{NM}$ and $\alpha_{FM}$ is given by [11, 22]

$$NLSV\ signal = \frac{2\alpha_{FM}^2 R_{NM}}{\left(2+\frac{R_{NM}}{R_{FM}}\right)^2 e^{d/\lambda_{NM}} - \left(\frac{R_{NM}}{R_{FM}}\right)^2 e^{-d/\lambda_{NM}}} \quad , \tag{1}$$

where FM= Py, NM = Cu or Ag, $R_{NM} = 2\lambda_{NM}/\rho_{NM}S_{NM}$ and $R_{FM} = 2\lambda_{FM}/\rho_{FM}S_{FM}(1-\alpha_{FM}^2)$ are spin resistances, $\lambda_{FM,NM}$ spin diffusion lengths, $\rho_{FM,NM}$ resistivities, and $S_{FM,NM}$ cross-sectional areas of FM and NM. For all samples, we use $\lambda_{Py}$= 5 nm [12, 19, 23] and $\rho_{Py}$ = 19 μΩcm. $\rho_{Py}$ was measured on a separate device deposited under nominally identical conditions, and is in agreement with values reported in the literature.[12, 19, 24] All other variables, $\rho_{Cu,Ag}$, $S_{Py}$, $S_{Cu,Ag}$ and *d*, were measured explicitly for each device. Finally, the data from each sample is fitted to Eq. 1 using $\alpha_{Py}$ and $\lambda_{Cu,Ag}$ as fitting parameters. We note that $\alpha_{Py}$ extracted in this way includes interface scattering and thermal effects and thus is not exactly the same as the intrinsic spin polarization of Py.[13]

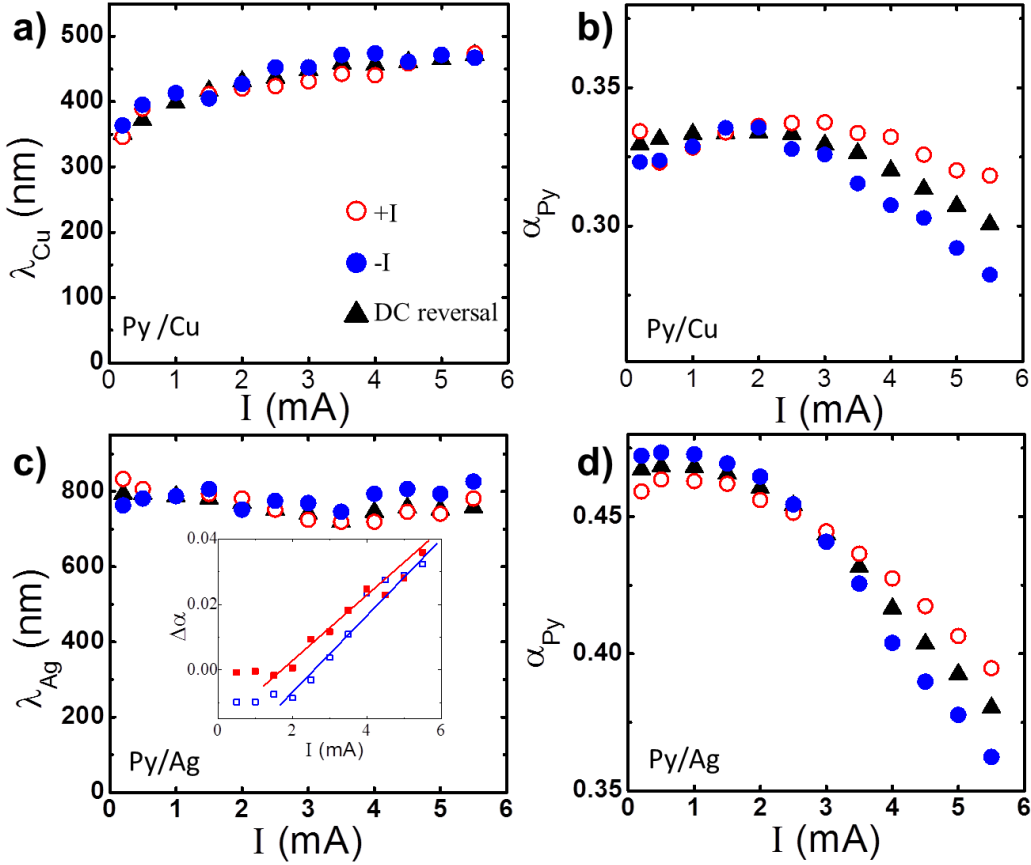

FIG. 3. **(a) and (b)** Py/Cu device. Cu spin diffusion length (a) and Py spin polarization (b) vs. current amplitude for positive currents (red empty circles), negative currents (blue full circles) and dc-reversal (black triangles). **(c) and (d)** Same for Py/Ag device. **Inset**: Difference between spin polarization of positive and negative currents vs. current magnitude. Red full squares correspond to Py/Cu system and blue empty squares to Py/Ag.

Figure 3 summarizes the results of the fits for the spin diffusion length and the effective spin polarization as a function of magnitude and polarity of the driving currents for the Py/Cu ($\lambda_{Cu}$ in Fig. 3a and $\alpha_{Py}$ in 3b) and for the Py/Ag ($\lambda_{Ag}$ in Fig. 3c and $\alpha_{Py}$ in 3d) devices. The Cu and Ag spin diffusion lengths depend on the current magnitude but *not* on its direction. The effective spin polarization, on the other hand, has strong current-polarity dependence (Fig. 3b and 3d for Py/Cu and Py/Ag, respectively). Both types of devices show a similar trend. At low currents (up to ~ 2.5 mA) the spin injection efficiencies are equal for positive and negative currents. On the other hand, above this current, $\alpha_{Py}$ is larger for positive than negative currents. The difference increases with increasing the current up to the maximum applied current ( ±5.5 mA). The difference in effective spin polarization, $\Delta\alpha_{Py} = \alpha_{Py}^{+I} - \alpha_{Py}^{-I}$, vs. current is plotted in the inset of Fig. 3c, for both the Py/Cu and Py/Ag devices. Both show that above a current threshold there is a linear dependence of $\Delta\alpha_{Py}$ on $I$. This indicates that the effect probably has the same origin within the Py electrode.

Since $\alpha_{Py}$ is the ratio between the spin current inside the Py, $J_S$, and the charge current $J_C$ ($J_S = \alpha_{Py} J_C$) in the Py, the observed linear dependence between $\Delta\alpha_{Py}$ and $I$ (shown in Fig. 3c, inset) indicates that the difference in injected spins is proportional to $I^2$ ($\Delta J_S \sim \Delta\alpha_{Py} J_C \sim J_C^2$). This suggests that the origin of the asymmetry is related to Joule heating.

An intriguing possibility is that 'spin blockade' [25] gives raise to the asymmetry; which is similar to developing a depletion layer at the copper interface. The current density where spin blockade is expected to appear can be estimated using Eq. (11) in Ref. [26]. Using for the Cu electron density $n=8.5\times10^{22}$ cm$^{-3}$, Fermi velocity $v_F = 1.57\times10^6$ m/s [27] and from our measurements $\lambda_{Cu}=400$ nm and $\rho_{Cu}=0.04$ $\Omega\mu$m; the approximated current density is $\sim10^{14}$ A/m$^2$. This value is three orders of magnitude higher than what was used in our measurements, and is above the critical brake-down current for our devices.

A realistic explanation for the asymmetry observed here may arise from the FM spin-dependent Seebeck effect, [9] in which a temperature gradient $\nabla T$ in the FM produces a spin current along this gradient. This can act as a spin source in specially designed lateral spin valves.[9] We found that this explains our results in a natural way; the sign of the effect, the dependence of the parameters as a function of injection current and gives the right order of magnitude for the observations.

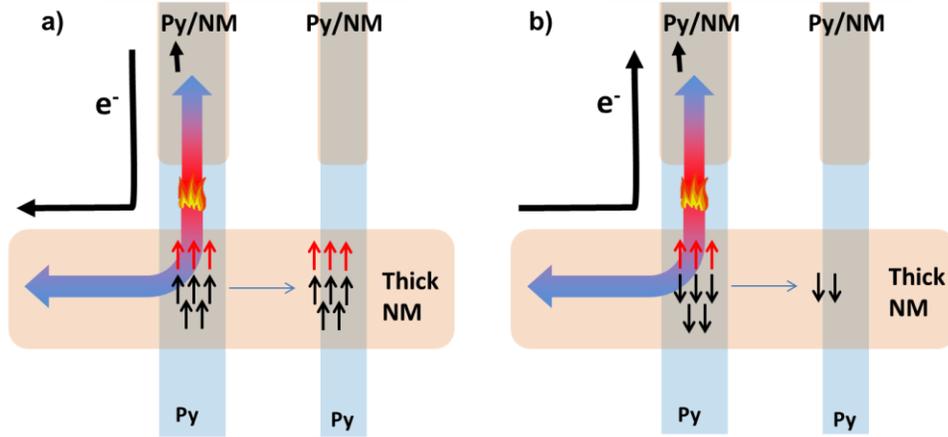

FIG. 4. Cartoon model of both the current-induced spin injection (black spins) and thermally-induced spin injection (red spins). For positive current (**a**) the up-spins combine to provide a larger signal, while negative currents (**b**) result in a smaller spin accumulation. The detecting electrode signal is proportional to the injection.

Here, most of the Joule heating arises from the current flowing in the high resistance bare part of the FM electrode (see cartoon in Fig. 4). This induces a large temperature gradient, $\nabla T$, in the FM near the FM/NM interface. Moreover, $\nabla T$ is the same for positive and negative currents since it is proportional to $I^2$. Consequently, the *thermal* injection (spin-dependent Seebeck effect) produces the same spin population for both current directions. However, changing the charge current direction results in *electrical* injection of opposite spin populations into the NM channel (i.e. injection vs. extraction).[11] Because of the negative sign of the Py spin-dependent Seebeck coefficient,[9] for positive current the effective spin injection ($J_S^+$) is the sum of the electrical spin injection and the thermally induced spin injection ($J_{SS}$): $J_S^+ = \alpha_{FM}J_C + J_{SS}$. Similarly, the effective spin injection for negative currents ($J_S^-$) is the difference of the electrical spin extraction and the thermal spin injection: $J_S^- = \alpha_{FM}J_C - J_{SS}$. This is in agreement with our measurements in which the NLSV signal for positive current injection is larger than for the negative current.

In addition to producing the correct sign, thermal spin injection is also in agreement with the measurements presented in Figure 3. Thermal spin injection modifies only the effective spin polarization ($\alpha_{Py}$), and not the spin diffusion length ($\lambda_{Cu,Ag}$), explaining the similarities between the Py/Cu and Py/Ag devices as shown in Fig. 3. Finally, the difference between the spin injection for positive and negative currents is proportional to the spin-dependent Seebeck induced injection, which is proportional to the temperature gradient. We can estimate the temperature gradient and spin-dependent Seebeck coefficient in our measurements using the expression for the thermally-induced spin current in the FM [9]:

$$J_{SS} = -\frac{(1-\alpha_{FM}^2)S_S^{FM}\nabla T}{2\rho_{FM}}. \quad (2)$$

Here $\nabla T$ is the temperature gradient along the FM close to the FM/NM injection interface and $S_S^{FM}$ is the spin-dependent Seebeck coefficient of the FM. From Eq. 2 we can estimate the difference in effective spin polarization for opposite current directions ($\Delta\alpha_{Py}$), using the expression for total spin-current injected for positive ($J_S^+$) and negative ($J_S^-$) charge-currents: $J_S^\pm = \alpha_{Py}J_C \pm J_{SS} = J_C(\alpha_{Py} \pm J_{SS}/J_C) = J_C(\alpha_{Py} \pm \Delta\alpha_{Py}/2)$:

$$\Delta\alpha_{Py} = \frac{(1-\alpha_{Py}^2)S_S^{Py}\nabla T}{\rho_{Py}J_C}. \quad (3)$$

Since $\nabla T$ is a function of the local Joule heating, which is proportional to $I^2$, we recover the linear dependence between $\Delta\alpha_{Py}$ and $J_C$ shown in the inset of Fig 3c. Using the literature spin-dependent Seebeck coefficient for Py, $S_S^{Py}= -3.8$ μVK$^{-1}$,[9] we find for the Py/Cu device at 5 mA a temperature gradient near the injection interface $\nabla T$ ~100 K/μm. This is on the same order of the 50 K/μm gradient reported in Ref. [9]. Such a gradient is reasonable here, recalling that the current density is 9.5×10$^{11}$A/m$^2$ (just below device failure), the heating is mainly in the small bare Py electrode and the thick NM electrodes act as an efficient heat sink. Alternatively, this could indicate that the spin-dependent Seebeck coefficient of our Py is larger. Assuming a 50 K/μm temperature gradient we obtain $S_S^{Py}= -7.6$ μVK$^{-1}$.

In summary, we measured the dependence of effective spin injection and spin diffusion length on the polarity and magnitude of the injection current. We find an additional large thermally-induced spin injection at high currents due to a temperature gradient in the Py electrode. At high current densities (above 4×10$^{11}$A/m$^2$), the thermal spin injection accounts for about 12% of the total spin injection. This additional thermally induced spin injection arises from the spin-dependent Seebeck current and may affect the behavior of spintronic devices.

We thank J. Bass and Y. V. Pershin for useful discussions. Work supported by the Office of Basic Energy Science, U.S. Department of Energy, under grant DE FG03-87ER-45332. The authors acknowledge financial support from the Spanish MICINN (MAT2009-08494), the Basque Government (PI2011-1) and the European Commission (PIRG06-GA-2009-256470 and PIRG07-GA-2010-268357).